\documentclass[aps,pre,reprint,superscriptaddress,showpacs,notitlepage,a4]{revtex4-2}
\usepackage{appendix}
\usepackage{bm}
\usepackage[colorlinks=true, linkcolor=black, urlcolor=blue, citecolor=blue]{hyperref}
\usepackage{hyperref}
\usepackage{xcolor}
\usepackage{comment}
\usepackage{geometry}
\usepackage{soul}
\usepackage{multirow}
\usepackage{graphicx} 
\usepackage{amsmath}
\usepackage{enumitem}

\newcommand{\taumax}{\tau_\text{max}}
\newcommand{\Tc}{T_{\mathrm{c}}}
\newcommand{\TcMF}{T_{\mathrm{c}}^{\mathrm{MF}}}

\begin{document}

\title{Stochastic field effects in a two-state system: symmetry breaking and symmetry restoring}

\author{Sara Oliver-Bonafoux}
\affiliation{Instituto de Física Interdisciplinar y Sistemas Complejos IFISC (CSIC-UIB), Campus UIB, 07122 Palma de Mallorca, Spain}
\author{Ra\'ul Toral}
\affiliation{Instituto de Física Interdisciplinar y Sistemas Complejos IFISC (CSIC-UIB), Campus UIB, 07122 Palma de Mallorca, Spain}
\author{Amitabha Chakrabarti}
\affiliation{Department of Physics, Kansas State University, Manhattan, KS 66506, USA}

\date{\today}

\begin{abstract}
We study the Ising model under a time-varying, but spatially homogeneous, Gaussian random magnetic field. In the Monte Carlo simulations, we go beyond the standard analysis of the order parameter by measuring the magnetization probability distribution as a function of temperature and field strength, and by computing the time required for the system to escape from a completely ordered state of the magnetization. We identify three distinct phases: a broad-paramagnetic phase, a broad-ferromagnetic phase and a bona-fide ferromagnetic phase. These broad phases display wide magnetization distributions that tend to limiting forms that remain finite in both height and width in the thermodynamic limit. The transition between the broad-paramagnetic and broad-ferromagnetic phases is a noise-induced transition and, for small field amplitudes, occurs at the critical temperature of the field-free Ising model. The transition from the broad-ferromagnetic to the ferromagnetic phase occurs at lower temperatures and is discontinuous, yet it does not fall into the conventional first-order class. Instead, it is characterized by a diverging escape time from an ordered magnetization state.
\end{abstract}

\maketitle

\section{Introduction}

The study of the Ising model subject to time-dependent, but spatially homogeneous, random magnetic fields has attracted considerable attention in recent decades. To characterize the phase behavior of the system, most studies focus on the order parameter $Q$, defined as the time-averaged magnetization, and examine its dependence on field strength and temperature. For a sinusoidally oscillating magnetic field~\cite{tome1990dynamic, lo1990ising, acharyya1995response, acharyya1999nonequilibrium}, the system undergoes a dynamic phase transition from a ferromagnetic phase ($Q \ne 0$) to a paramagnetic phase ($Q = 0$) as the temperature or field amplitude increases. This transition is continuous for low field amplitudes and becomes discontinuous at higher amplitudes, resulting in a tricritical point in the field-amplitude and temperature plane. Other studies, in contrast, consider a uniformly distributed, time-varying random magnetic field~\cite{Acharyya1998}, and report that the transition between the ordered (\makebox{$Q \ne 0$}) and disordered ($Q = 0$) phases remains continuous for all field amplitudes and temperatures.

In this work, we examine the Ising model under the presence of a time-dependent, but uniform in space, zero-mean Gaussian random magnetic field. We argue that a study based solely on the order parameter overlooks essential features of the system. A more complete picture emerges from analyzing the probability distribution of the magnetization across different system sizes, which provides deeper insight into the phase diagram of the system. 

We find several effects associated with the presence of the time-varying magnetic field. First, we show that a vanishing order parameter (\makebox{$Q = 0$}) may correspond to two distinct phases: a broad-paramagnetic phase and a broad-ferromagnetic phase. In the \textit{broad} phases, the magnetization displays wide distributions around their maxima values, and their width and height tend to finite values in the thermodynamic limit of an infinite system size. Furthermore, the broad-ferromagnetic phase is characterized by a dynamical restoring of the symmetry between the two equivalent ferromagnetic states~\cite{BPT:1994,BPTK:1997,Toral:2011}. Thus, the random magnetic field converts the conventional symmetry-breaking phase transition of the Ising model into a noise-induced transition---{\`a la Horsthemke and Lefever}~\cite{1984Horsthemke}---between the broad-paramagnetic and broad-ferromagnetic phases. For sufficiently small amplitudes of the field, this transition occurs at the same critical temperature of the field-free Ising model.

Second, at lower temperatures, we uncover a new kind of phase transition between the broad-ferromagnetic phase and a \textit{bona-fide} ferromagnetic phase, where one of the two equivalent states is selected dynamically and therefore the order parameter is non-zero ($Q \ne 0$). Although this transition is discontinuous, it does not fall within the conventional classification of first-order phase transitions. Instead, it is characterized by a divergence of the characteristic switching time between ordered states of the magnetization.

These findings are not only of theoretical interest but also have direct relevance for experimental studies. In the context of electrodeposition, classical crystal growth and electrochemical crystallization, the introduction of stochastic electric fields has been shown to enhance ion heating and reaction rates~\cite{martinez2013effective, paneru2023bona}. Such systems naturally involve the interplay of thermal noise and stochastic external driving, where stochastic resonance and related phenomena play a central role~\cite{Flanders, gammaitoni1998stochastic}. Our model sheds light on such experimental studies and we believe it will inspire further work on barrier hopping in two-state systems influenced by stochastic fields.

The paper is structured as follows: In Sec.~\ref{sec:Model}, we describe the model and the Monte Carlo simulation scheme. In Sec.~\ref{sec:Results}, we present the results: in Sec.~\ref{sec:MC_PhaseIdentification} we identify the three phases exhibited by the model, and in Sec.~\ref{sec:MC_TransitionsAnalysis} we discuss the transitions between them. In Sec.~\ref{sec:MF}, we report numerical results obtained within the mean-field description, which align with those observed in the two-dimensional case. Finally, our conclusions and a discussion of our results can be found in Sec.~\ref{sec:Conclusions}.

\section{Model and simulation scheme}\label{sec:Model}

Let us consider an Ising model in the presence of a time-varying, but spatially homogeneous, random magnetic field~\cite{Acharyya1998}. The Hamiltonian of the system is
\begin{equation}
 \mathcal{H} = - J \sum_{\langle i,j \rangle} s_i s_j - h(t) \sum_i s_i,
\end{equation}
where $s_i \in \{-1, +1\}$ is the spin variable, $J > 0$ is the strength of the ferromagnetic exchange between nearest-neighbor spins $\langle i,j\rangle$ and $h(t)$ is the time-dependent, but uniform in space, random magnetic field. We take $h(t)$ to be Gaussian distributed with zero mean and variance~$2D$, where the parameter $D$ sets the strength of the field. The $N = L \times L$ spins are distributed on a square lattice of side $L$ with periodic boundary conditions. 

Modeling the system in contact with an isothermal bath at temperature $T$, we introduce a heat bath stochastic dynamics with single spin random updates~\cite{Newman2023}. After setting randomly the initial configuration, the evolution is such that a randomly chosen spin adopts a new value $s_i=\pm1$, independent of the previous one, according to the probabilities
\begin{align}\label{eq:probabilities}
P^t(s_i = +1) &= \left(1 + e^{-2 \beta b_i}\right)^{-1}, \nonumber \\ 
P^t(s_i = -1) &= 1 - P^t(s_i = +1), 
\end{align}
with $\beta = 1 / (k T)$ and $b_i = J \sum_\mu s_{i_\mu} + h(t)$, where the sum over $\mu$ runs over the four nearest neighbors of the regular lattice. We have added a superscript $t$ to this probability to highlight its time dependence through the magnetic field $h(t)$. For simplicity, we set the Boltzmann constant, $k$, and the coupling constant,~$J$, equal to~$1$ throughout the paper. Time $t$ is measured in Monte Carlo steps (MCS), so that one unit of time corresponds to $N$ spin updates. The random variable $h(t)$ remains constant during a time interval $P$, after which a new random value $h(t+1)$ is selected independent of the previous one. Throughout this study, we fix \makebox{$P = 1$ MCS}.

\section{Results}\label{sec:Results}

\subsection{Phase identification}\label{sec:MC_PhaseIdentification}

First, we aim to identify the phases exhibited by the model in the presence of the random magnetic field and to contrast them with those of the field-free case ($D = 0$). For this purpose, we compute the probability distribution of the magnetization as a function of temperature $T$ and field strength $D$. In the simulations, after discarding an initial thermalization transient, we take measurements of the instantaneous magnetization,
\begin{equation}
 m(t) = \frac{1}{N} \sum_{i=1}^N s_i(t),
\end{equation}
and determine its time-averaged stationary probability distribution, $P(m)$. We limit our analysis to moderate values of field strength~$D$ to allow for a balanced competition between phase ordering and stochastic effects. For large values of \makebox{$D$ ($\gtrsim$ 1),} we observe a very rich phenomenology that lies beyond the scope of this work.

\begin{figure}[ht!]
\centering{\includegraphics[width=1.0\linewidth]{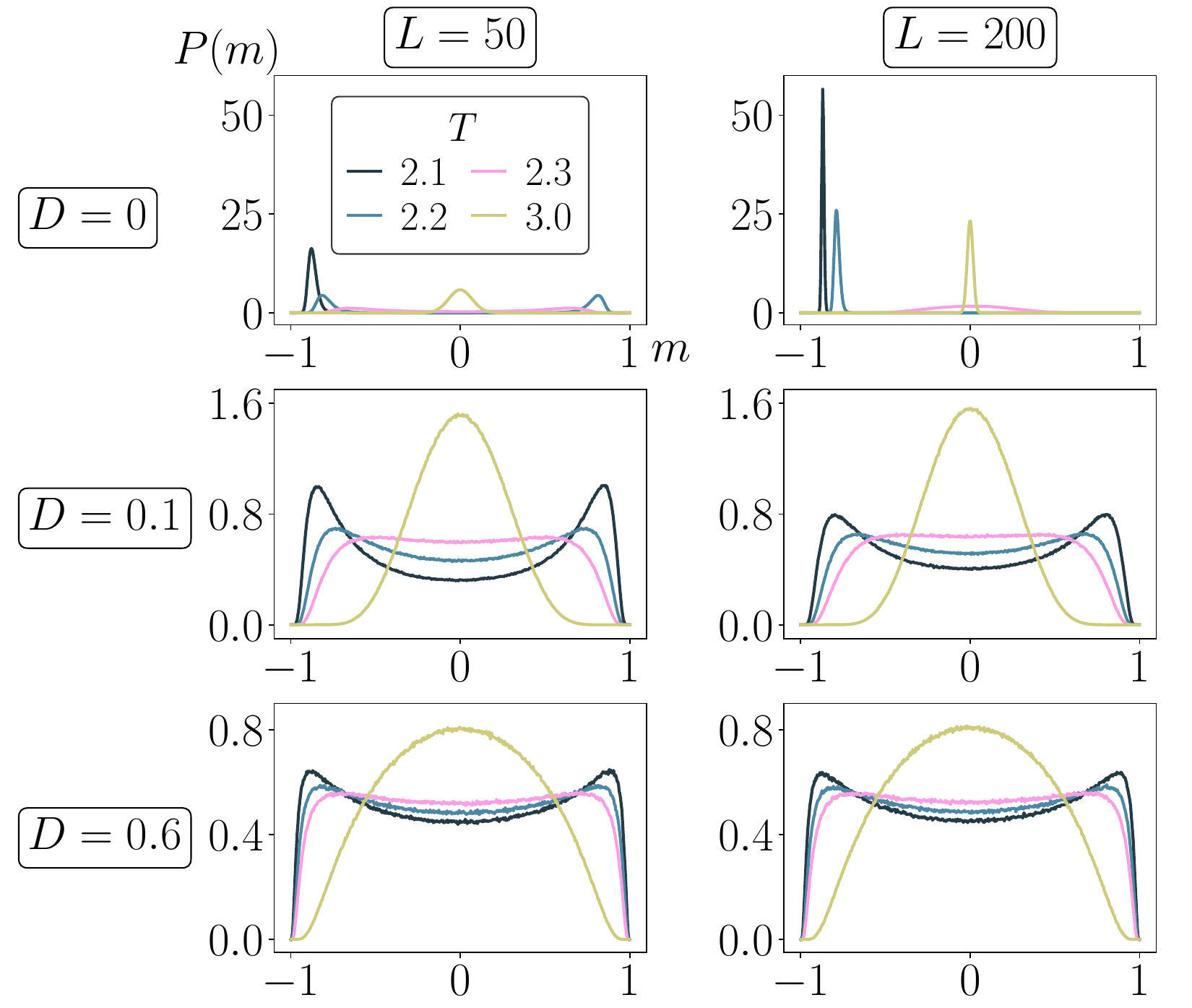}} 
 \caption{Broad-paramagnetic and broad-~ferromagnetic phases. Histograms of the normalized probability distribution of the magnetization, $P(m)$, for different field intensities ($D = 0, 0.1, 0.6$) and for two system sizes ($L = 50, 200$) at several temperatures. Note that graphs in the same row have the same vertical scale. For each set of parameters ($D, L, T$), results are obtained from a single agent-based numerical simulation. After a thermalization time of $10^6$ MCS, we let the system evolve for $10^8$ MCS and measure the instantaneous magnetization every 10 MCS. Thus, the distributions $P(m)$ are derived from a sample of $10^7$ measurements of $m(t)$. These and subsequent histograms presented in this work have been computed using 500 bins of width $\Delta m=0.004$ in the interval $m\in[-1,1]$. }\label{fig:HistogramsComparison}
\end{figure}

\subsubsection{Results in the field-free case ($D = 0$)}

Fig.~\ref{fig:HistogramsComparison} shows the probability distribution of the magnetization, $P(m)$, for three values of field strength~$D$ (including \makebox{$D = 0$}), at temperatures close to the two-dimensional Ising equilibrium critical temperature, \makebox{$\Tc \approx 2.269$}~\cite{onsager1944crystal}, and for two system sizes: $L = 50$ and $L = 200$. We first focus on the well-known Ising case, corresponding to $D=0$.

At temperatures below $\Tc$, such as $T=2.1$, the distribution exhibits a single peak at a non-zero magnetization value that gets sharper with increasing system size. Note that the two ordered magnetization states are equally probable. The results shown in Fig.~1 correspond to a single, long simulation run in which the system happens to order around $m\approx-1$; other realizations would yield a peak around $m\approx+1$ with the same probability. At temperature $T= 2.2$, which is close to, but still below $\Tc$, the distribution for the smallest system size ($L = 50$) shows two symmetric peaks close to $m \approx \pm 1$, indicating that the trajectories $m(t)$ are able to jump frequently between the two ordered states. These transitions, however, are absent for the largest system ($L = 200$). Above the critical temperature, i.e., for $T = 2.3$ and $T = 3.0$, the distribution exhibits a single peak at $m = 0$ that also becomes sharper as~$L$ increases.

All these observations align with the expected behavior in the limit of infinite system size: a ferromagnetic phase, characterized by a symmetry-breaking Dirac-delta distribution centered in one of the two ordered states, for $T < \Tc$; and a paramagnetic phase, characterized by a Dirac-delta distribution centered in the disordered state, for $T > \Tc$. 

\subsubsection{Broad-ferromagnetic and broad-paramagnetic phases}

We now turn to the results obtained for finite values of field strength~$D$. As can be seen in Fig.~\ref{fig:HistogramsComparison}, the system still undergoes a transition from a unimodal to a bimodal regime at the zero-field critical temperature, $\Tc$, as the temperature decreases. However, the displayed behavior is significantly different from the $D=0$ case. 

\begin{figure}[ht!]
 \centering{ \includegraphics[width=1.0\linewidth]{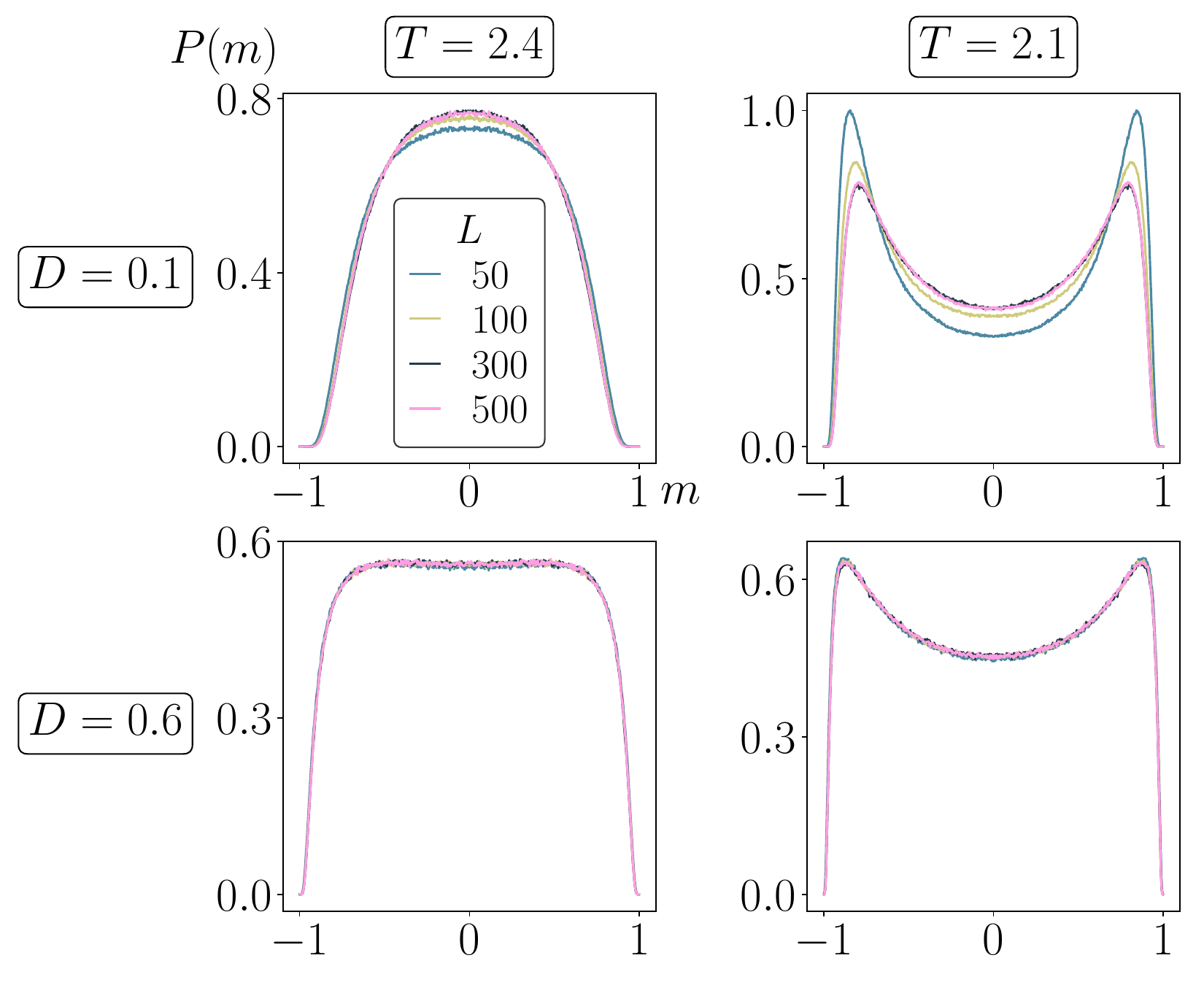}} \caption{Limiting wide distributions in the broad phases. Probability distribution of the magnetization, $P(m)$, for two field intensities (\makebox{$D = 0.1$ and $D = 0.6$}) at temperatures $T = 2.4$ and $T = 2.1$, which lie within the broad-paramagnetic and broad-ferromagnetic phases, respectively. Results are shown for four system sizes: $L = 50, 100, 300$ and 500. In the numerical simulations, we have taken $10^7$ measurements of the instantaneous magnetization after an initial transient of $10^6$ MCS. Note that, for $D = 0.1$, the distributions for $L=300$ and $L=500$ overlap at the scale of the figure for both temperatures.}\label{fig:LimitingDistribution}
\end{figure}

Notably, both for $D = 0.1$ and $D = 0.6$, at temperature $T = 2.0$ (below $\Tc$), even for the largest system size ($L = 200$), the dynamical trajectories display frequent jumps between the two ordered states of magnetization. On the other hand, at temperature $T = 3.0$ (above $\Tc$), the probability distribution for the largest system ($L = 200$) is still centered around $m = 0$, but is much wider than in the $D = 0$ scenario. Additionally, the $P(m)$ distributions show much less dependence on system size than in the field-free case, suggesting that a wide distribution is obtained in the thermodynamic limit. Note that the smallest system ($L = 50$) exhibits sharper distributions than the largest system ($L = 200$) at a fixed temperature. This phenomenon is illustrated more clearly in Fig.~\ref{fig:LimitingDistribution}, which shows the probability distribution of the magnetization for $D = 0.1$ and $D = 0.6$ at temperatures $T = 2.4$ and $T = 2.1$, corresponding to the broad-paramagnetic and broad-ferromagnetic phases, respectively. In the figure we consider larger system sizes $L$ than those shown in Fig.~\ref{fig:HistogramsComparison}. The distributions for large $L$ completely overlap, indicating that a limiting wide distribution is reached as the system size increases.

Consequently, in the limit $L \to \infty$, we expect the following behavior: for $T\lesssim \Tc$, a \textit{broad-ferromagnetic} phase where the symmetry between the two ordered magnetization states is dynamically restored, leading to a probability distribution with two broad peaks centered around these states; and for $T > \Tc$, a \textit{broad-paramagnetic} phase characterized by a broad probability distribution around the disordered value of magnetization.

\subsubsection{Ferromagnetic phase}

\begin{figure}[ht!]
 \centering{ \includegraphics[width=1.0\linewidth]{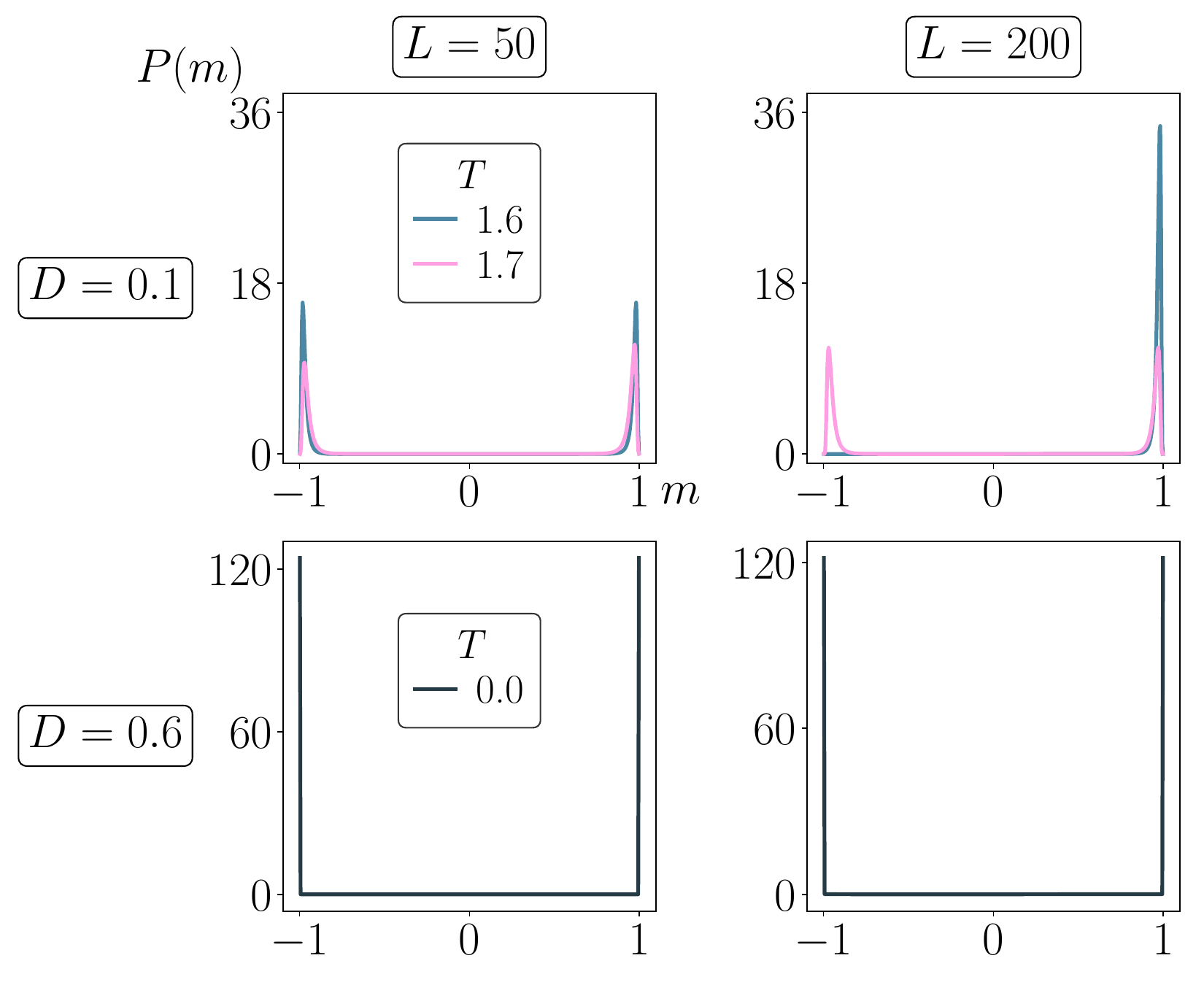}} \caption{Ferromagnetic phase. Probability distribution of the magnetization, $P(m)$, for two system sizes ($L = 50, 200$) and for two field intensities ($D = 0.1, 0.6$) at different temperatures. Note that graphs in the same row have the same vertical scale. In the numerical simulations, the thermalization and measurement times have been set identical to those in Fig.~\ref{fig:HistogramsComparison}.}\label{fig:HistogramsD01D06}
\end{figure}

Let us now examine the case $D>0$ at temperatures significantly below the critical point. Fig.~\ref{fig:HistogramsD01D06} shows the probability distribution of magnetization, $P(m)$, for field intensities $D = 0.1$ and $D = 0.6$ at lower temperatures than those presented in Fig.~\ref{fig:HistogramsComparison}. 

We first focus on the results obtained for \makebox{$D = 0.1$}. At temperature $T = 1.7$, the independence of the distributions with respect to system size, along with the observed dynamical restoring between the two ordered magnetization states, indicates that the system is in the broad-ferromagnetic phase. In contrast, at a slightly lower temperature ($T = 1.6$), the displayed behavior resembles that of the field-free Ising model: the symmetry restoring phenomenon is a finite-size effect that is only exhibited by the smallest system ($L = 50$), whereas for the largest system ($L = 200$) the symmetry is broken and the system dynamically selects one of the two equivalent stable states. These results suggest that, at this temperature, the system is in a \textit{bona-fide} ferromagnetic phase. However, it is still possible that, over a longer observation time (exceeding the $10^8$ MCS considered in Fig.~\ref{fig:HistogramsD01D06}), a dynamic switching between the two ordered magnetization states could occur across all system sizes. Such behavior would indicate that the system might actually fall within the broad-ferromagnetic phase. Therefore, what we can conclude is that for $D = 0.1$ and an observation time of $10^8$ MCS, a transition between the broad-ferromagnetic and ferromagnetic phases occurs within the temperature range $1.6 < T < 1.7$. The nature of this transition and its dependence on observation time will be analyzed later. 

We now discuss the results obtained for a field strength $D = 0.6$. In this case, due to the enhanced effects of the random magnetic field, the broad-ferromagnetic phase is exhibited even at temperature $T = 0.0$ and the ferromagnetic phase is not observed at any temperature.

\subsubsection{Region of validity of the different phases}

\begin{figure}[ht!]
 \centering{ \includegraphics[width=0.75\linewidth]{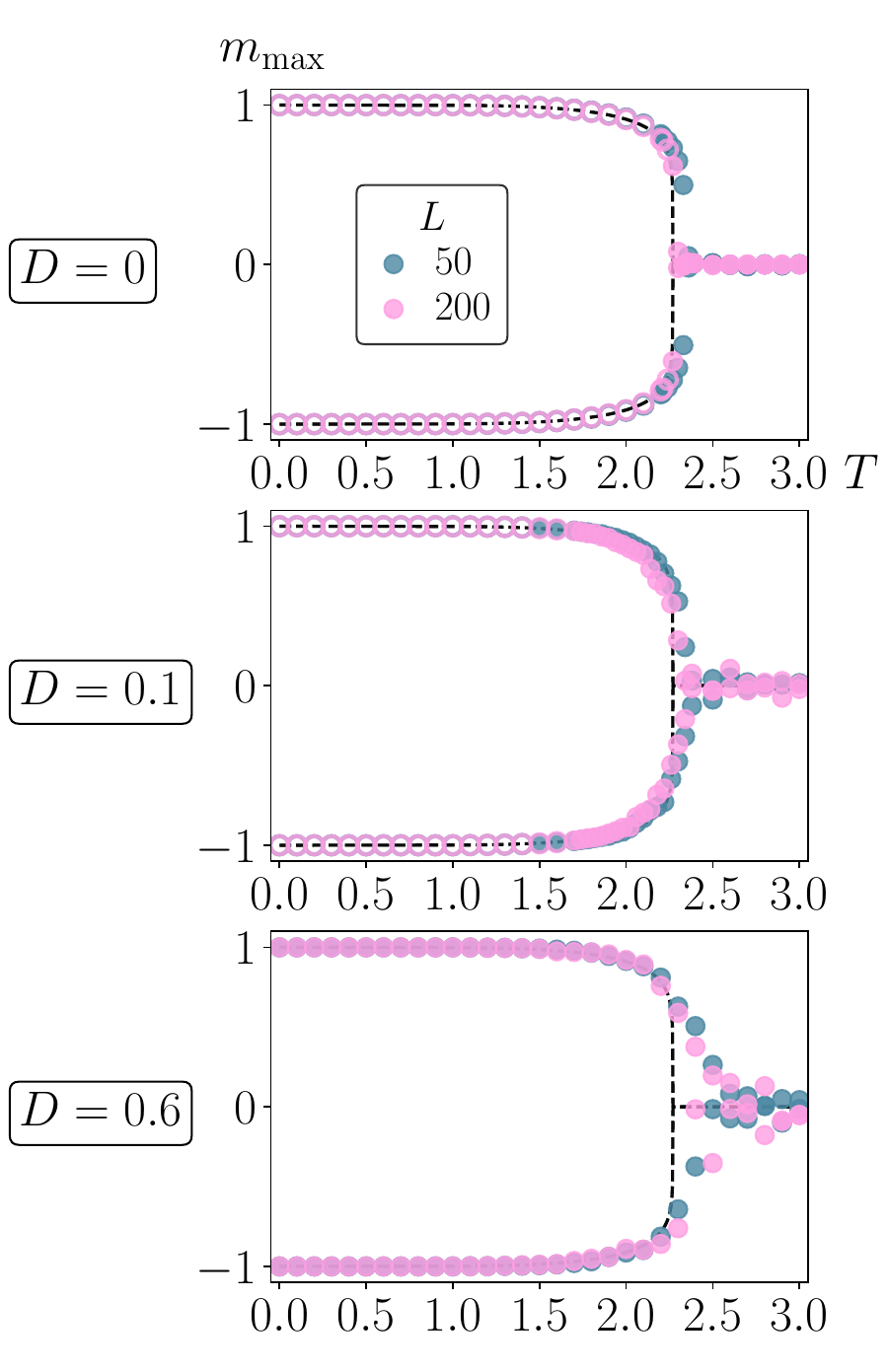}}
\caption{Region of validity of the different phases. Maxima $m_\mathrm{max}$ of the probability distribution of magnetization, $P(m)$, as a function of temperature for several intensities of the random magnetic field ($D = 0$, 0.1, 0.6) and for two system sizes ($L = 50, 200$). The black dashed lines correspond to the Onsager's theoretical solution, while symbols represent the outcomes from numerical simulations. Open circles represent the ferromagnetic phase, while colored circles correspond either to the broad-ferromagnetic phase (two circles located at two symmetric values of~$m$) or to the paramagnetic phase---true or broad---(a single circle around $m = 0$). The same thermalization and measurement times as in Fig.~\ref{fig:HistogramsComparison} have been applied here.}\label{fig:mmaxvsT}
\end{figure}

Fig.~\ref{fig:mmaxvsT} provides a comprehensive overview of the different phases exhibited in the model. The figure illustrates the maxima of the probability distribution of magnetization, $m_\mathrm{max}$, as a function of temperature for different field intensities, and for two system sizes. The Onsager's theoretical solution~\cite{Baxter2011} for the field-free Ising model is also depicted in the graphs. The three distinct phases that the system can exhibit are represented as follows: open circles located at two symmetric values, in the case of a ferromagnetic phase with symmetry breaking; colored circles, still symmetrically located, in the case of a broad-ferromagnetic phase with symmetry restoring; or colored circles around $m = 0$, in the case of a paramagnetic phase (including the broad case).

For $D = 0$, the exhibition of the broad-ferromagnetic phase, a bimodal phase with symmetry restoring, is a finite-size effect that disappears as the system size increases. In contrast, for both $D = 0.1$ and $D = 0.6$, the broad-ferromagnetic phase is exhibited, within a certain temperature range, for all system sizes. The temperature range in which the system is in this phase becomes larger as the field intensity increases. For the observation time considered in the figure ($10^8$ MCS), this temperature range corresponds to $1.6 \lesssim T \lesssim \Tc$ for $D = 0.1$ and to $0 \leq T \lesssim \Tc$ for $D = 0.6$. These findings suggest that there is a critical value of the field strength,~$D_\mathrm{c}$, at which the ferromagnetic phase vanishes and the bimodal phase is always accompanied by a dynamical restoring of the symmetry between the two ordered states of magnetization.

Finally, we note that for $D > 0$ the curves also fit Onsager's theoretical solution. Notably, for \makebox{$D = 0.1$}, and in contrast to the $D = 0$ scenario, even the results for the smallest system size ($L = 50$) match more closely the theoretical solution than those for the largest system ($L = 200$).

\subsection{Analysis of the transitions}\label{sec:MC_TransitionsAnalysis}

\subsubsection{Phase diagram}

\begin{figure}[ht!]
 \centering{ \includegraphics[width=1\linewidth]{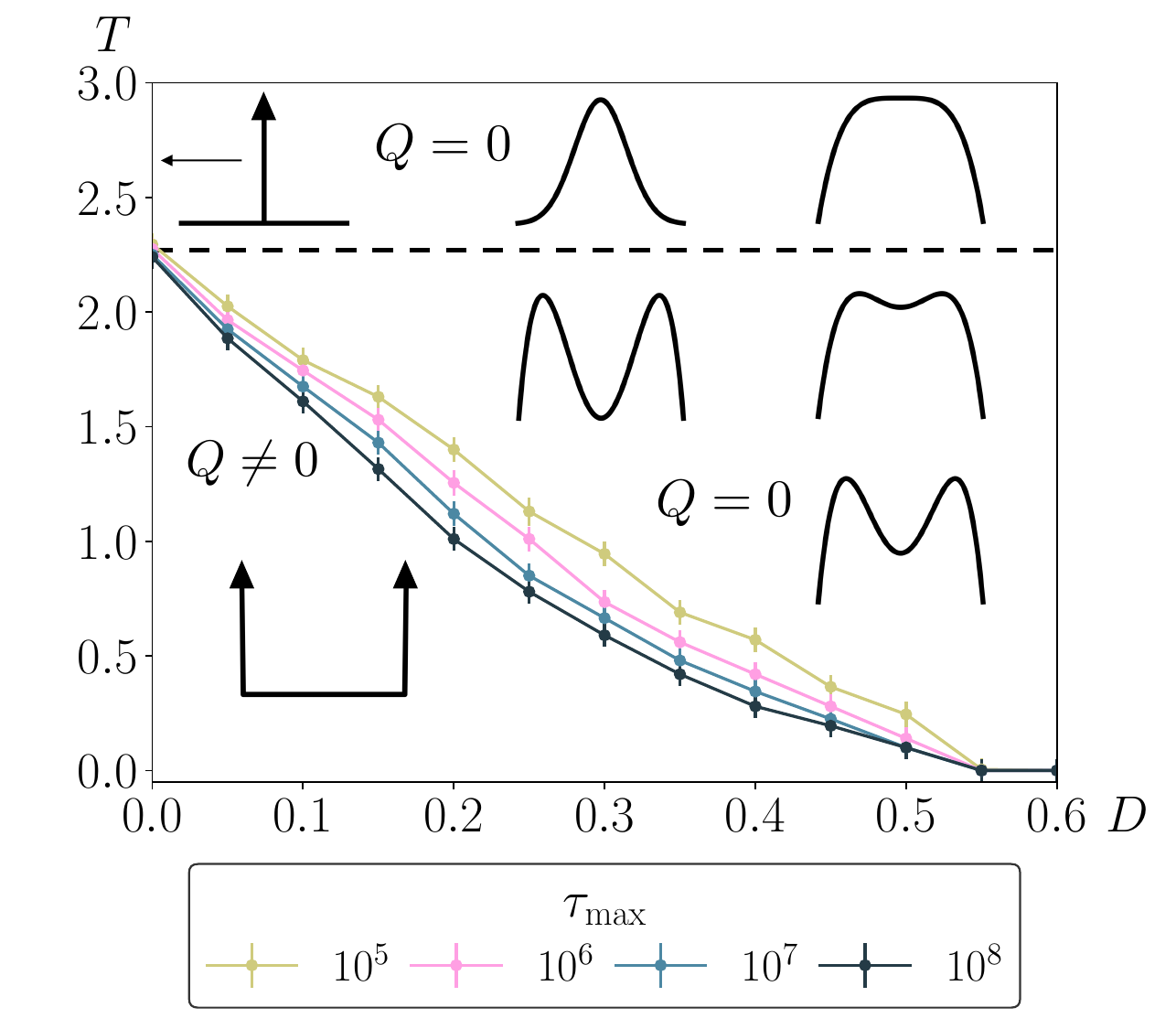}}
 \caption{Phase diagram in the $(D,T)$ plane. The diagram indicates the expected shape of the probability distribution of the magnetization, together with the corresponding order parameter, in the thermodynamic limit for each region. The arrows in the probability distribution indicate a Dirac-delta contribution. The horizontal dashed line corresponds to the critical temperature of the field-free Ising model, $\Tc \approx 2.269$. The colored lines separating the broad-ferromagnetic and ferromagnetic phases come from numerical simulations for a system of size $L = 100$. In the simulations, given a value of $D$, we look for the first temperature at which the system is able to jump from $m = -1$ to $m = 0$ in a time shorter than a given observation time $\taumax$. We then identify this temperature with the temperature of separation between the broad-ferromagnetic and ferromagnetic phases for that combination of $D$ and $\taumax$ values. The curves in the figure represent the average of 10 repetitions of this procedure.}\label{fig:PhaseDiagram}
\end{figure}

The above results lead us to characterize the numerical phase diagram of the model in the $(D,T)$ plane, as shown in Fig.~\ref{fig:PhaseDiagram}. The diagram illustrates the characteristic shape of the magnetization probability distribution, along with the corresponding order parameter, expected in the thermodynamic limit in the different regions of the diagram.

At the critical temperature~$\Tc$, it is well established that a symmetry-breaking phase transition occurs between the paramagnetic and ferromagnetic phases in the field-free case. Due to the addition of the random magnetic field, this phase transition transforms into a noise-induced transition ~\cite{1984Horsthemke}, which is not accompanied by symmetry breaking in the thermodynamic limit. This transition, which occurs at the same critical temperature $\Tc$ of the field-free model, separates the broad-paramagnetic and broad-ferromagnetic phases. 

At a temperature below $\Tc$, the presence of the random field induces a new phase transition between the broad-ferromagnetic and ferromagnetic phases. This phase transition occurs at a temperature that decreases as the field strength increases, and disappears for field intensities \makebox{$D > D_\mathrm{c} \approx 0.6$}.

The order parameter $Q$, defined as the time-averaged magnetization, can be obtained as \makebox{$Q = (1/T) \sum_{t=T_0}^{T_0+T} m(t)$}, where $T_0$ and $T$ denote the thermalization and measurement times, respectively. Equivalently, this quantity can be computed from the probability distribution of the magnetization as
\begin{equation}\label{eq:OrderParameter}
Q = \sum_m m P(m).
\end{equation}
As shown in Fig.~\ref{fig:PhaseDiagram}, the order parameter vanishes in both the broad-paramagnetic and broad-ferromagnetic phases, while it is non-zero in the ferromagnetic phase.

\subsubsection{Characteristic jump time}

In order to identify the boundary in the phase diagram separating the broad-ferromagnetic and ferromagnetic phases, we introduce the characteristic disordering time $\tau$. We define this quantity arbitrarily, but precisely, as the average time the system takes to reach the disordered state $m = 0$ starting from the completely ordered state \makebox{$m = -1$}.

\begin{figure*}[ht!]
 \centering{\includegraphics[width=0.8\linewidth]{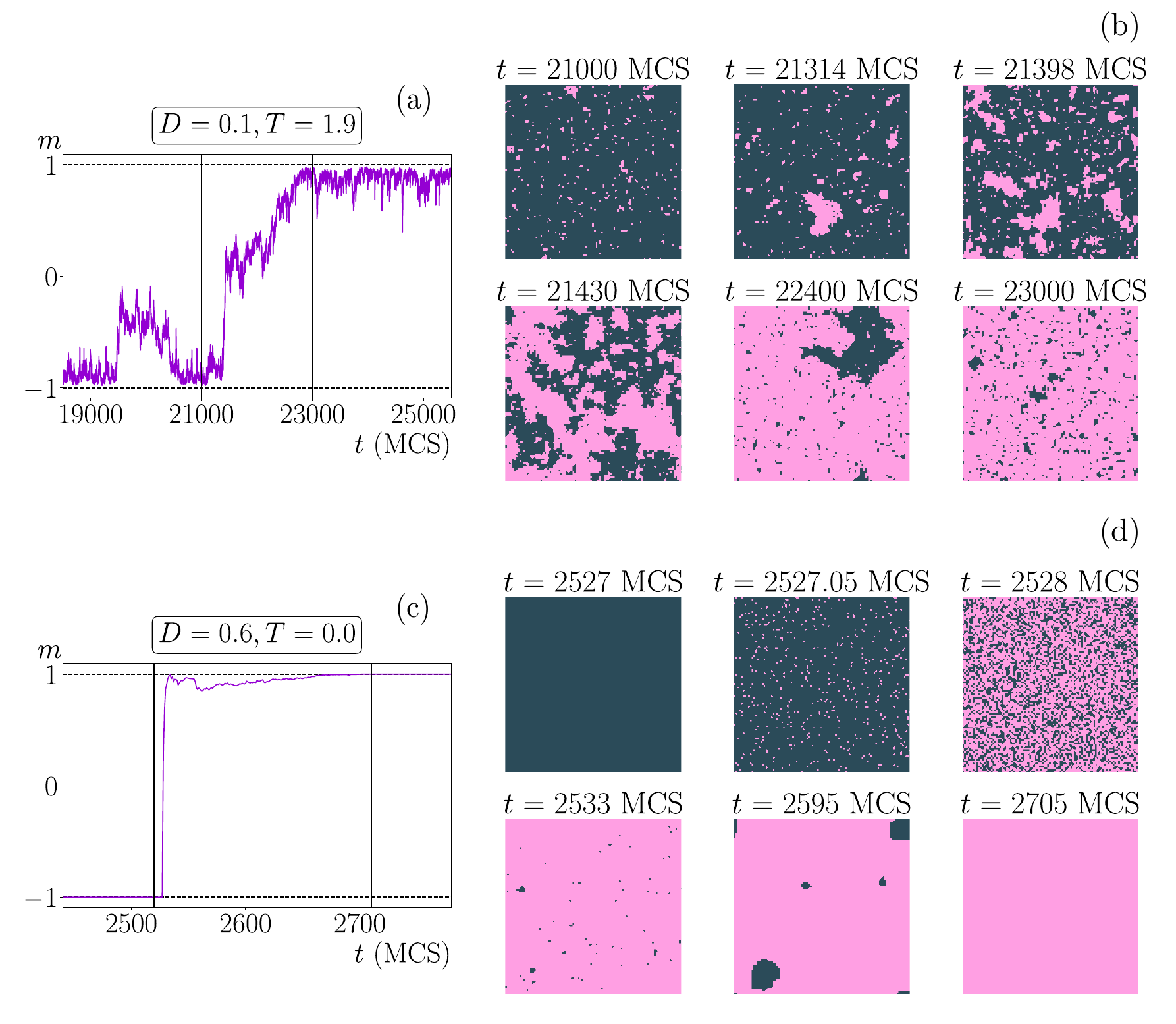}} \caption{Jumps between ordered states in the broad-ferromagnetic phase. Left panels: trajectories of the magnetization for a system of size $L = 100$. Panel (a) corresponds to $D = 0.1$ and $T = 1.9$, and panel (c) to $D = 0.6$ and $T = 0.0$. Both combinations of parameters $(D,T)$ lie within the broad-ferromagnetic phase. In both cases, the system is initialized in the state $m(t=0)=-1$. The vertical dashed lines indicate the time intervals during which the system jumps from $m \simeq -1$ to $m \simeq 1$: (a) $t = 21000$--$23000$ MCS, and (c) $t = 2520$--$2710$ MCS. Right panels: selected snapshots of the spatial spin configuration corresponding to the trajectories shown in (a) and (c), displayed in panels (b) and (d), respectively. Spins in state $-1$ are shown in gray, while spins in state $+1$ are shown in pink. Movies of the complete transitions are publicly available in the GitHub repository~\cite{BroadFerroSnapshots}.}\label {fig:TrajectorySnapshots}
\end{figure*}

As discussed above, the system is able to jump from one ordered state to the other in the broad-ferromagnetic phase. In Fig.~\ref{fig:TrajectorySnapshots}, we show two characteristic examples of such transitions for different combinations of field intensity $D$ and temperature~$T$ (panels (a) and (c)). Panels (b) and (d) display selected snapshots of the corresponding spatial spin configurations taken along the trajectories shown in (a) and (c), respectively. For $D = 0.1$ and $T = 1.9$, the transition occurs via the nucleation and growth of droplets of the new phase. For $D = 0.6$ and $T = 0.0$, the change of phase is initiated by the flipping of isolated spins, leading to a rapid change in the global magnetization. Droplets of the original phase may still form and persist for some time, but they eventually disappear and the system becomes fully aligned in the opposite state.

These examples illustrate that, in the broad-ferromagnetic phase, a system  starting in the completely ordered state $m = -1$ crosses the barrier $m = 0$ after a finite time on its way to the opposite ordered state, $m = +1$. This behavior is expected to occur even in the limit of infinite system size. In the ferromagnetic phase, in contrast, the system is not able to escape from an ordered state, resulting in a jump time than tends to infinity as the system size increases.

The colored lines in Fig.~\ref{fig:PhaseDiagram} separate, for a system of size $L = 100$, the region where the system can jump from $m = -1$ to $m = 0$ in a time less than $\taumax$ (which corresponds to the broad-ferromagnetic phase), from the region in which the system is not able to make the transition (which corresponds to the ferromagnetic phase). As the observation time $\taumax$ increases, the area of the ferromagnetic region decreases. Numerically, we cannot rule out the possibility that this region vanishes in the limit $\taumax \to \infty$, but we believe this is an unlikely scenario. Our results indicate that, for all practical purposes, the system exhibits both the broad-ferromagnetic and ferromagnetic phases. We note here that, in experimental realizations of crystal growth phenomena, the time scale over which the field fluctuates (typically MHz frequency~\cite{Flanders}) is much shorter than the typical observation time.

\subsubsection{Classification of the transition between the broad-ferromagnetic and ferromagnetic phases}

\begin{figure}[ht!]
 \centering{ \includegraphics[width=1.0\linewidth]{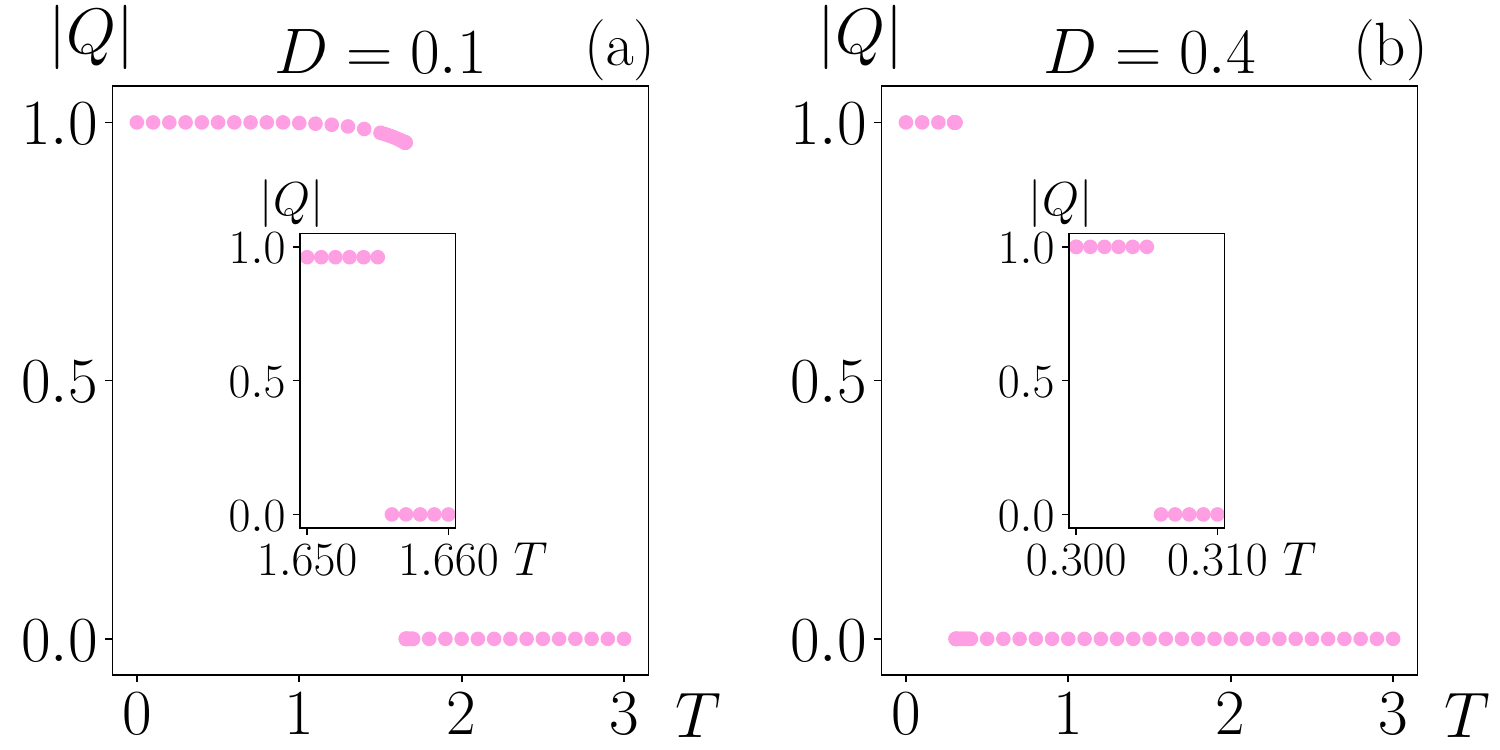}} \caption{Absolute value of the order parameter as a function of temperature for a system of size $L = 200$ and a field strength: (a) $D = 0.1$ and (b) $D = 0.4$. For each temperature $T$, we initialize the system with a random configuration and let it evolve. After discarding a thermalization transient, we compute the probability distribution of the magnetization and determine the order parameter using Eq.~\eqref{eq:OrderParameter}. The thermalization and measurement times are the same as in Fig.~\ref{fig:HistogramsComparison}. Insets: enlargement of the curves for (a) $1.650 \le T \le 1.660$ and (b) $0.300 \le T \le 0.310$.}\label{fig:OrderParameter}
\end{figure}

\begin{figure}[ht!]
 \centering{ \includegraphics[width=0.7\linewidth]{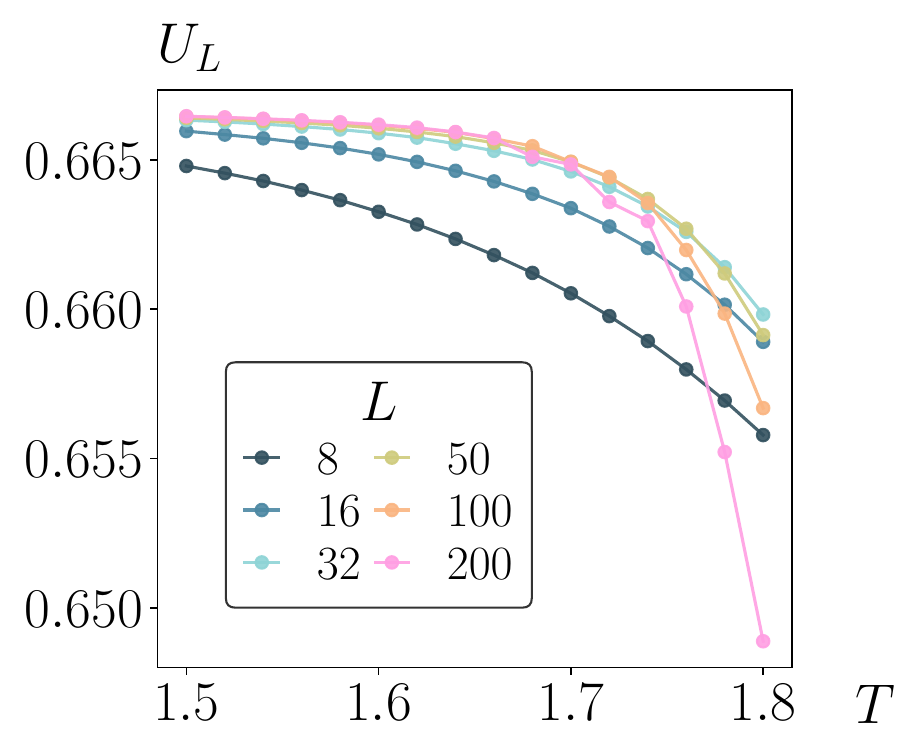}} \caption{Fourth-order Binder cumulant, defined as \makebox{$U_L = 1 - \langle m^4 \rangle / (3 \langle m^2 \rangle^2)$}, as a function of temperature for a field intensity $D = 0.1$ and several system sizes. The $k$th moment of the magnetization is defined as $\langle m^k \rangle = \sum_m m^k P(m)$. The data comes from numerical simulations with the same thermalization and measurement times as in Fig.~\ref{fig:HistogramsComparison}.}\label{fig:BinderCumulant}
\end{figure}

In Fig.~\ref{fig:OrderParameter}, we show the absolute value of the order parameter, $|Q|$, as a function of temperature for a system of size $L = 200$ and for field intensities \makebox{$D = 0.1$} and $D = 0.4$ (panels (a) and (b), respectively). As the temperature decreases, the order parameter exhibits an abrupt change from zero to a non-zero value in the transition region between the broad-ferromagnetic and ferromagnetic phases, indicating that the transition is discontinuous for all field strengths $D$. Although this behavior could be interpreted as indicative of a first-order transition, the transition does not fall within the standard classification of first-order phase transitions. In particular, given the shapes of the probability distributions involved in the transition, there is no possibility of coexistence between ordered and disordered states (see Fig.~\ref{fig:PhaseDiagram}). Moreover, for large system sizes the fourth-order cumulant $U_L$ does not not display the pronounced negative minimum near the transition temperature that is characteristic of first-order transitions~\cite{vollmayr1993finite} (see Fig.~\ref{fig:BinderCumulant}).

A second-order transition is also ruled out, since the finite-size scaling analysis of the fourth-order Binder cumulant (see Fig.~\ref{fig:BinderCumulant}) does not reveal a crossing of the curves corresponding to different system sizes near the transition temperature~\cite{landau2021guide}. Therefore, the calculation of critical exponents is not meaningful in the model with an external random field.

Unlike conventional first- or second-order, this transition between a broad-ferromagnetic and a ferromagnetic phase exhibits a different hallmark of critical behavior. This transition is characterized by the divergence of the jump time~$\tau$, which quantifies the timescale associated with escaping from a fully ordered state of magnetization.

\section{Mean-field analysis}\label{sec:MF}

In this section, we present numerical results obtained within the mean-field description of the model, which reproduce the qualitative behavior observed in the two-dimensional agent-based model.

As detailed in Sec.~\ref{sec:Model}, the system evolves through single spin updates, where a randomly chosen spin adopts a new state independent of the previous one. In the mean-field description, the probability in Eq.~\eqref{eq:probabilities} for the new state to be $s_i = +1$ can be approximated as
\begin{equation}
P^t(s_i = +1) \approx \left\{1 + e^{-\frac{2}{T} \left[\frac{8 n(t)}{N} - 4 + h(t)\right]}\right\}^{-1},
\end{equation}
where $n(t) = [m(t)+1] N / 2$ denotes the number of spins in state $+1$ at time $t$. 

After the update of spin $i$, i.e., after a time \makebox{$\Delta t = 1 / N$}, the three possible values for $n(t + \Delta t)$ are: $n(t) + 1$, $n(t) - 1$, or $n(t)$, depending on whether spin $i$ flips from $-1$ to $+1$, from $+1$ to~$-1$, or remains unchanged. Denoting the probability $P\left[n(t+\Delta t) = n' | n(t) = n\right]$ as $P^t(n \to n')$, the corresponding probabilities are 
\begin{align}\label{eq:MF_probabilities}
\begin{split}
&P^t\left(n \to n + 1\right) = \tfrac{N-n}{N}\, P^t(s_i = +1), \\
&P^t\left(n \to n - 1\right) = \tfrac{n}{N}\,\bigl[1 - P^t(s_i = +1)\bigr], \\
&P^t\left(n \to n \right) = 1 - P^t\left(n \to n + 1\right) - P^t\left(n \to n - 1\right),
\end{split}
\end{align}
where we have used that the fractions of spins in states +1 and $-1$ are $n / N$ and \makebox{$(N-n)/N$}, respectively. 

According to the transition probabilities in Eq.~\eqref{eq:MF_probabilities}, the Chapman--Kolmogorov equation governing the time evolution of the probability distribution of the process reads
\begin{align}\label{eq:MF_ChapmanKolmogorov}
P(n,t+\Delta t) &= P(n,t)\,P^t(n\to n) \nonumber \\
&\quad + P(n+1,t)\,P^t(n+1\to n) \nonumber \\
&\quad + P(n-1,t)\,P^t(n-1\to n),
\end{align}
which is valid for $n = 0, 1, \dots, N$.

The evolution equation can be written as \makebox{$\vec{P}(t+\Delta t)=W(t)\vec{P}(t)$}, where we have used the vector notation $\vec{P}(t)=(P(0,t),\dots,P(N,t))$ for the probability distribution, and $W(t)$ is a tridiagonal transition probability matrix with elements \makebox{$W_{n,m}(t)=P^t(m\to n)$}. This matrix remains constant during $N$ updating steps, before taking new random values, leading to $\vec{P}(t+1)=W(t)^N\vec{P}(t)$. Iterating from an initial condition $\vec{P}(0)$ we can write Eq.~\eqref{eq:MF_ChapmanKolmogorov} in the equivalent form
\begin{equation}\label{eq:MF_Matrix}
\vec{P}(t)=\prod_{k=0}^{t-1}W(k)^N\vec{P}(0),
\end{equation}
where each $W(k)$ is obtained using a different realization $h(k)$ of the Gaussian field.

\subsection{Probability distribution of the magnetization}\label{sec:MF_ProbabilityDistribution}

The probability distribution of the magnetization can be obtained by iterating numerically Eq.~\eqref{eq:MF_Matrix} up to a final time $T$ and discarding $T_0$ MC steps for thermalization. The time-averaged probability distribution is then estimated as
\begin{equation}
\vec{P} = \frac{1}{T} \sum_{t=T_0}^{T_0+T} \vec{P}(t).
\end{equation}
In the figures, we represent the probability distribution of the magnetization $m$, which is related to that of $n$ as $P(m) = P(n) \frac{N}{2}$.

Our choice of the initial distribution \makebox{$P(n,0)=\delta_{n,0}$}, which corresponds to a completely ordered state of the magnetization, does not prevent the system from evolving towards the broad-paramagnetic and broad-ferromagnetic phases, as discussed below.

\begin{figure}[ht!]
\centering{ \includegraphics[width=1.0\linewidth]{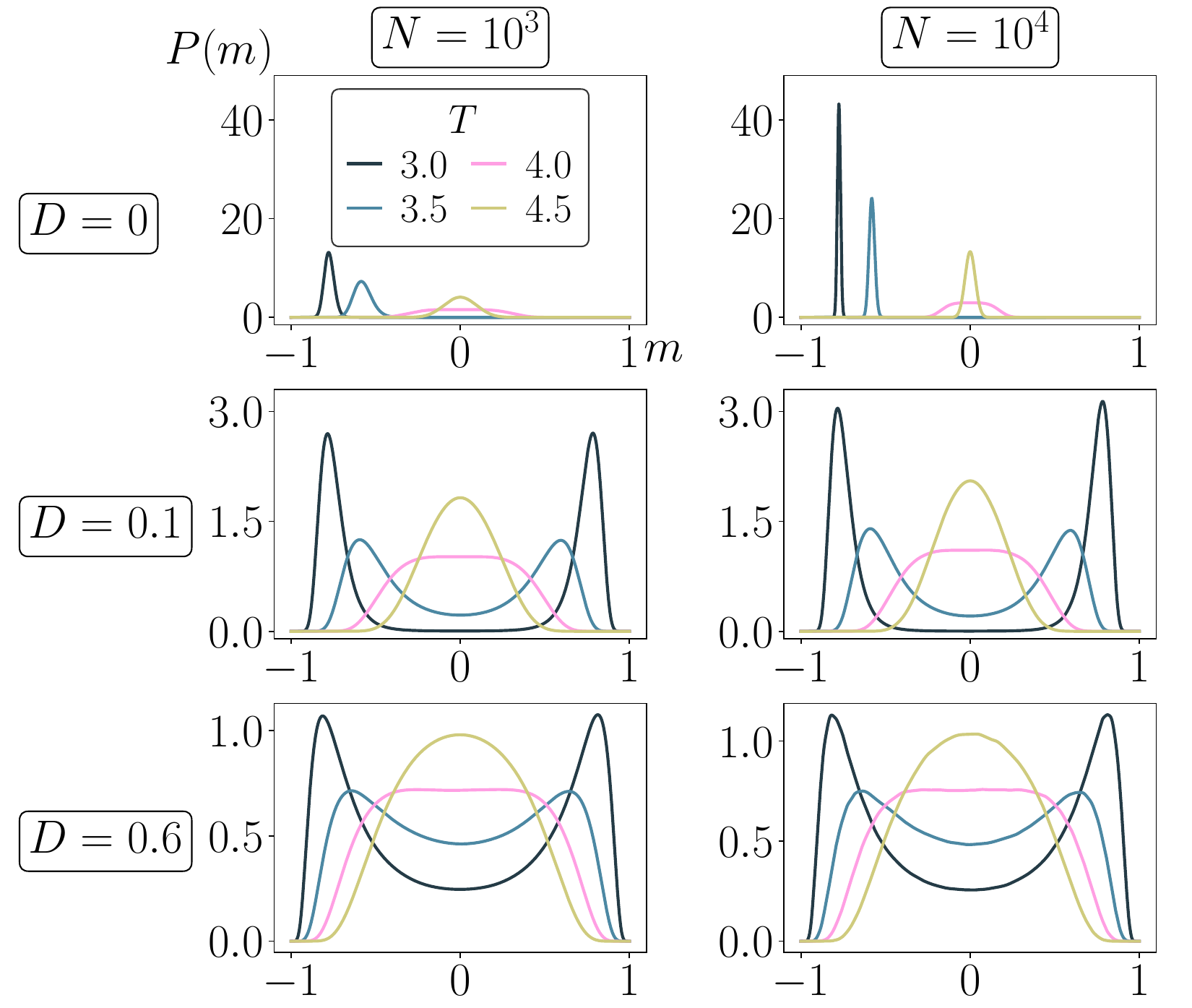}}
\caption{Broad-paramagnetic and broad-ferromagnetic phases in the mean-field description. Histograms of the normalized probability distribution of the magnetization, $P(m)$, for two system sizes (\makebox{$N = 10^3, 10^4$}) and for three field intensities ($D = 0, 0.1, 0.6$) at several temperatures. Note that graphs in the same row have the same vertical scale. The data comes from iterating Eq.~\eqref{eq:MF_ChapmanKolmogorov} as described in the text. We have set $T_0 = 10^3$ and $T = 10^4$ for $D = 0$; and $T_0 = 10^5$ and $T = 2 \cdot 10^6$ for $D = 0.1, 0.6$.}\label{fig:MF_HistogramComparison}
\end{figure}

In Fig.~\ref{fig:MF_HistogramComparison}, we show the probability distribution of the magnetization for three field intensities (\makebox{$D = 0, 0.1, 0.6$}) at several temperatures, and for two system sizes. First, we note that the critical temperature of the two-dimensional Ising model, \makebox{$\Tc \approx 2.269$}, shifts to $\TcMF = 4$ in the mean-field theoretical framework. This result, well-established in the field-free case~\cite{huang2008statistical}, also holds for field intensities $D > 0$. For $D = 0$, the system undergoes a phase transition between the paramagnetic and ferromagnetic phases at temperature $\TcMF$. Due to the addition of the random magnetic field, this phase transition becomes a noise-induced transition between a broad-paramagnetic phase and a broad-ferromagnetic phase. Both phases are characterized by broad magnetization distributions centered around their maxima values, with both width and height remaining finite in the limit of large size.

\begin{figure}[ht!]
\centering{ 
\includegraphics[width=1\linewidth]{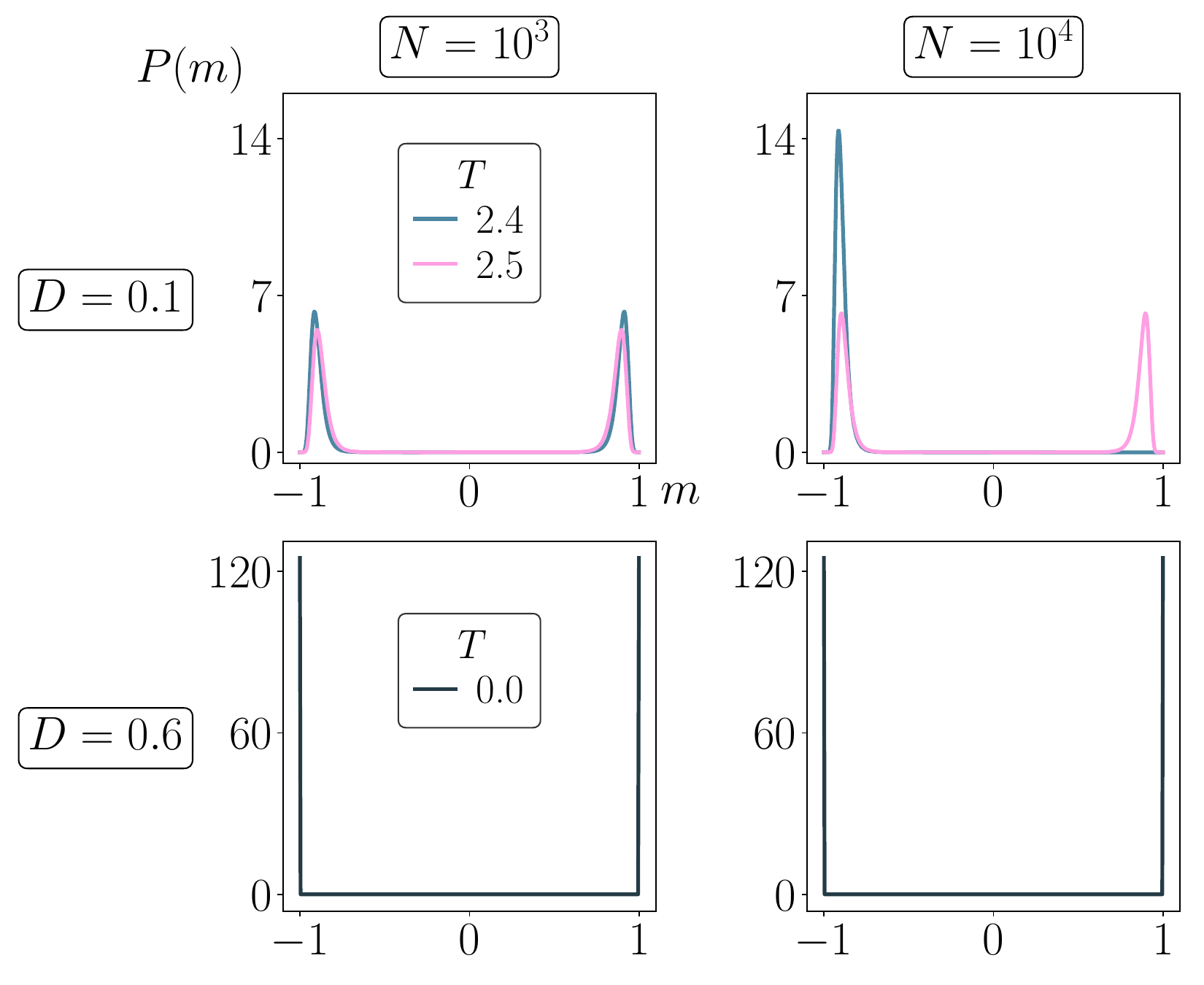}}
\caption{Ferromagnetic phase in the mean-field description. Probability distribution of the magnetization for two field intensities ($D = 0.1, 0.6$) and for two system sizes ($N = 10^3, 10^4$) at several temperatures. Note that graphs in the same row have the same vertical scale. In the numerical simulations, the thermalization and measurement times have been set identical to those in Fig.~\ref{fig:MF_HistogramComparison}.}\label{fig:MF_HistogramComparison2}
\end{figure}

In Fig.~\ref{fig:MF_HistogramComparison2}, we show the histograms of the magnetization for field intensities $D = 0.1$ and $D = 0.6$ at lower temperatures than those considered in Fig.~\ref{fig:MF_HistogramComparison}. For $D = 0.1$, and for the observation time considered in the figure, the system exhibits a transition between the broad-ferromagnetic phase and a bona-fide ferromagnetic phase within the temperature range $2.4 < T < 2.5$. For $D = 0.6$, in contrast, the bimodal phase is accompanied by symmetry restoring even at $T = 0.0$, and the ferromagnetic phase is thus not exhibited at any temperature. These results indicate that, for $D > 0$, the system undergoes a transition from the broad-ferromagnetic phase to the ferromagnetic phase at a temperature $T < \TcMF$. This temperature decreases with increasing field intensity $D$, and vanishes for sufficiently large $D$.

\begin{figure}[ht!]
 \centering{ \includegraphics[width=0.75\linewidth]{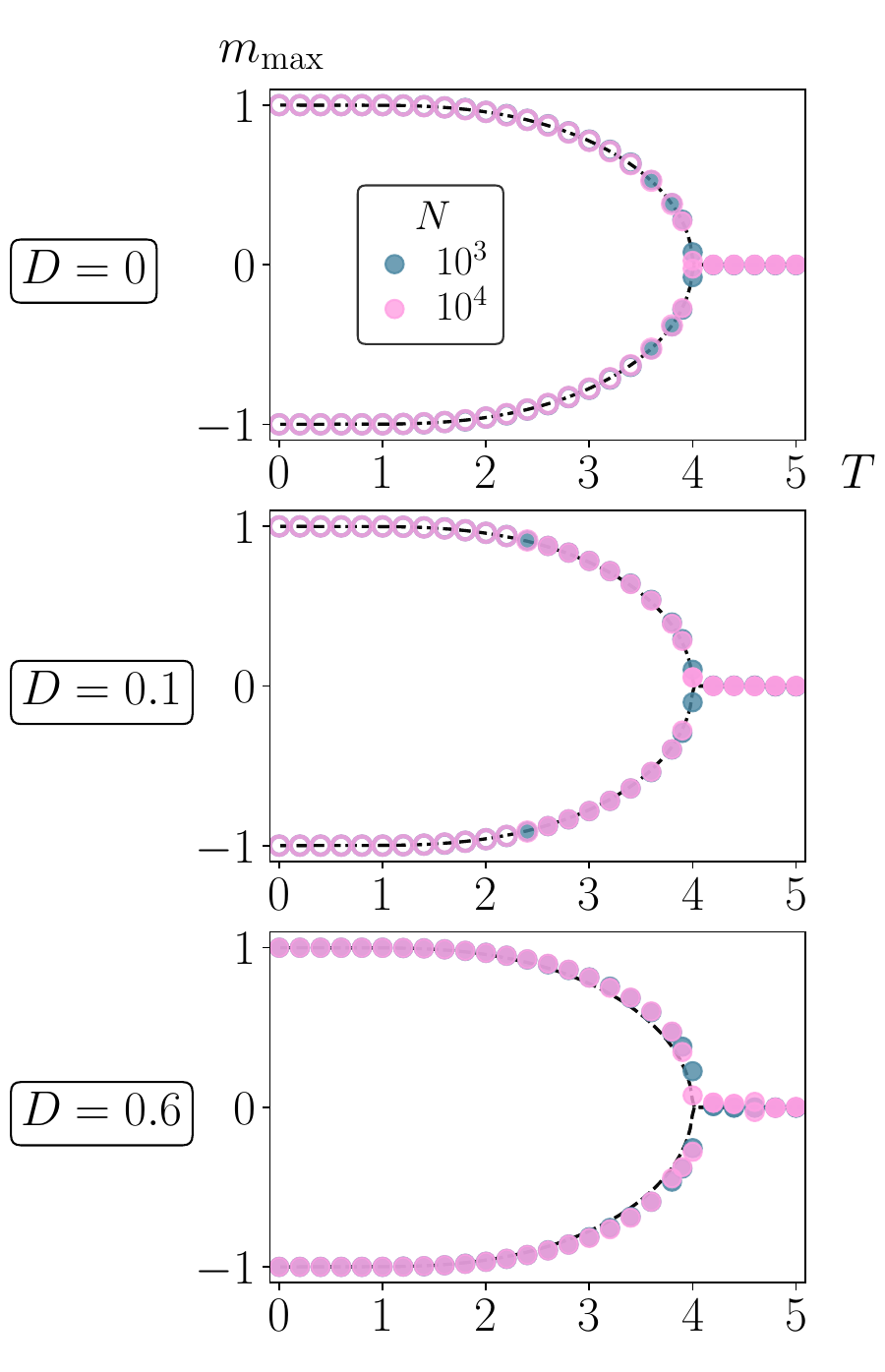}}
\caption{Region of validity of the different phases in the mean-field description. Maxima $m_\mathrm{max}$ of the probability distribution of magnetization, $P(m)$, as a function of temperature for several field strengths ($D = 0$, 0.1, 0.6) and for two system sizes ($N = 10^3, 10^4$). The black dashed lines correspond to the mean-field theoretical solution for $D = 0$ (obtained by numerically solving Eq.~\eqref{eq:MF_mmaxtheo}), while symbols represent the outcomes from numerical simulations. The symbol code used to represent the different phases is the same as that used in the two-dimensional case (see Fig.~\ref{fig:mmaxvsT}). In the simulations, we have applied the same thermalization and measurement times as in Fig.~\ref{fig:MF_HistogramComparison}.}\label{fig:MF_mmaxvsT}
\end{figure}

In Fig.~\ref{fig:MF_mmaxvsT}, we show the maxima of the probability distribution of magnetization, $m_\mathrm{max}$, as a function of temperature for several field intensities, and for two system sizes. To distinguish the different phases, we use the same symbol code as in the two-dimensional case (see Fig.~\ref{fig:mmaxvsT}). The black dashed lines represent the mean-field theoretical solution in the field-free case, which is obtained by numerically solving the equation~\cite{huang2008statistical}
\begin{equation}\label{eq:MF_mmaxtheo}
 m_\mathrm{max} = \tanh\left({\frac{4 \, m_\mathrm{max}}{T}}\right).
\end{equation}
As in the two-dimensional case, the region corresponding to the ferromagnetic phase decreases as the field intensity increases, eventually disappearing for large field values. Moreover, the curves obtained for $D > 0$ also agree with the theoretical solution obtained in the field-free case.

All these results indicate that the phase behavior observed in the mean-field description is consistent with that of the two-dimensional agent-based model.

\subsection{Separation boundary between the broad-ferromagnetic and ferromagnetic phases}\label{sec:MF_Boundary}

\begin{figure}[ht!]
\centering{ 
\includegraphics[width=1\linewidth]{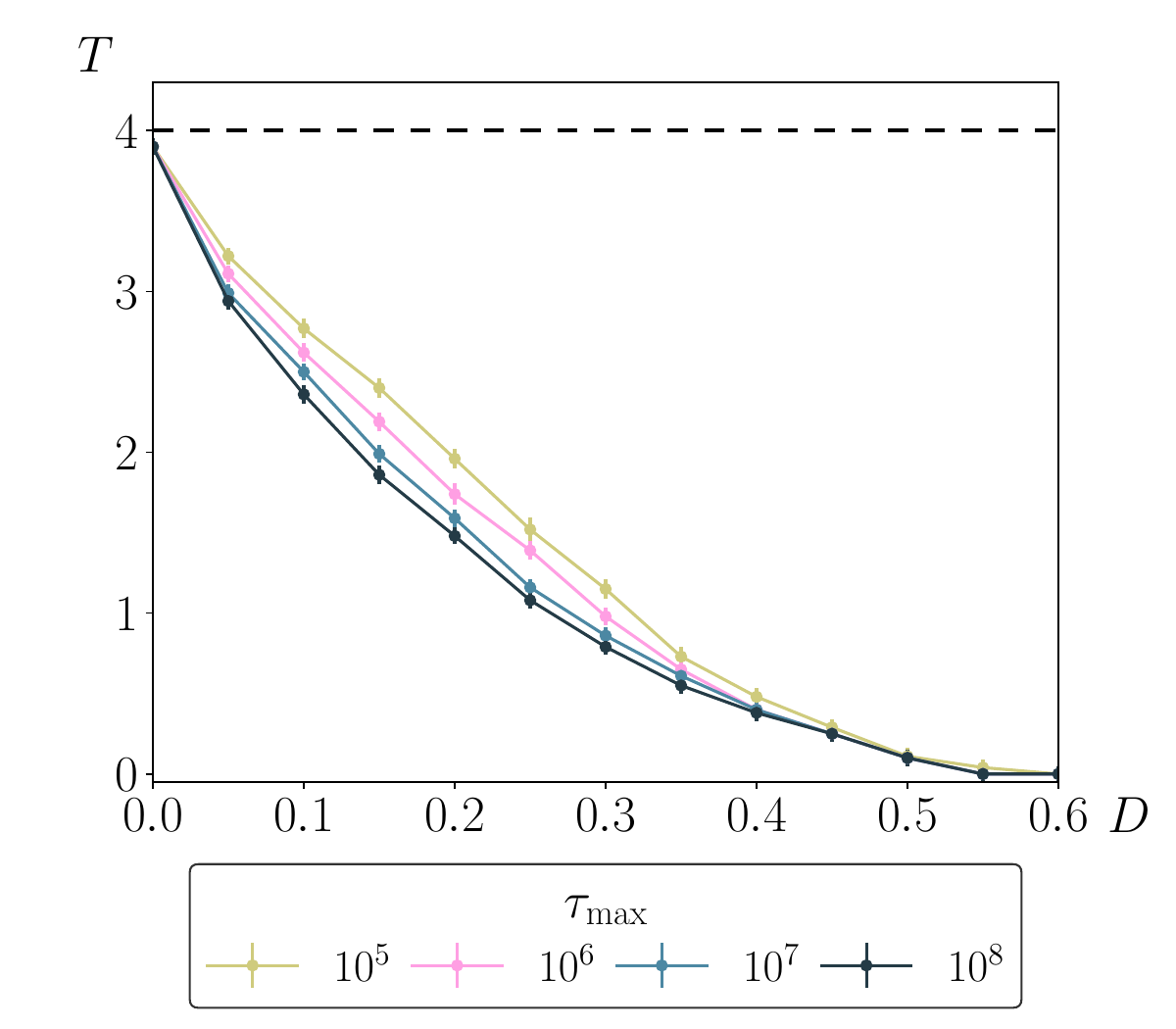}}
\caption{Separation boundary between the broad-ferromagnetic and ferromagnetic phases in the $(D, T)$ plane in the mean-field description. The colored lines, which separate the broad-ferromagnetic and ferromagnetic phases, have been obtained using the same criterion as in the two-dimensional agent-based problem (see Fig. 4) for a system of size \makebox{$N = 10^4$}. Specifically, for fixed values of $D$ and $\taumax$, the transition temperature is defined as the lowest temperature at which the system can transition from the completely ordered state $n = 0$ to the disordered state $n = N/2$. The dynamics of $n$ are simulated using the transition probabilities in Eq.~\eqref{eq:MF_probabilities}, where the magnetic field $h(t)$ is updated every MCS. Each result is averaged over a sample of 10 independent realizations of this procedure. The horizontal dashed line indicates the critical temperature of the field-free Ising model in the mean-field description, $\TcMF = 4$. 
}\label{fig:MF_PhaseDiagram}
\end{figure}

In Fig.~\ref{fig:MF_PhaseDiagram}, we show the boundary in the $(D, T)$ plane separating the broad-ferromagnetic and ferromagnetic phases for different observation times. The separation lines between the two phases have been obtained using the same method as in the two-dimensional model (see Fig.~\ref{fig:PhaseDiagram}), and correspond to a system of size $N = 10^4$. 

The mean-field analysis provides additional numerical evidence that, as the observation time $\taumax$ increases, the separation boundary approaches a limiting form rather than vanishing in the limit $\taumax \to \infty$. Moreover, the critical value of the field strength at which the ferromagnetic phase disappears and the broad-paramagnetic phase is exhibited at all temperatures $T < \TcMF$ is $D_\mathrm{c} \approx 0.6$, as in the two-dimensional case.

\section{Conclusions}\label{sec:Conclusions}

In this work, we have examined a two-dimensional Ising model in the presence of a time-varying, but spatially homogeneous, Gaussian random magnetic field. The study has been carried out using both Monte Carlo simulations of the two-dimensional model and a mean-field theoretical analysis, yielding consistent results across the two approaches.

While most prior research has focused on the analysis of the time-averaged magnetization, we have demonstrated that the probability distribution of magnetization provides deeper insight about the phase behavior of the system. Specifically, histograms of the magnetization are the key element for the identification of the broad-paramagnetic, broad-ferromagnetic and ferromagnetic phases in the model.

Using this approach, we have distinguished a broad-paramagnetic phase, a unimodal phase with a broad probability distribution around the disordered value of magnetization, and a broad-ferromagnetic phase, a bimodal phase exhibiting a dynamical restoring of the symmetry between the two equivalent ferromagnetic states. In the thermodynamic limit, both broad phases exhibit probability distributions that tend to limiting forms that remain finite in both height and width. The transition separating these phases is a noise-induced transition that, for small amplitudes of the magnetic field, occurs at the critical temperature of the field-free Ising model. These phases have not been reported in prior studies of related models, which primarily focused on the time-averaged magnetization (see, e.g.,~\cite{Acharyya1998}).

We have found that the phase transition separating the broad-ferromagnetic phase and the bona-fide ferromagnetic phase displays unusual features. While the transition is discontinuous, it does not fall within the conventional classification of first-order phase transitions. Instead, it is characterized by the divergence of the characteristic jump time~$\tau$, defined as the time the system takes to escape from a completely ordered state of the magnetization. Introducing this measure has allowed us to define the boundary in the phase diagram separating the two phases involved in the transition.

We believe that this work will illuminate theoretical foundations behind experimental studies and will lead to additional experiments of barrier hopping in two-state systems in the presence of stochastic fields. A possible extension of the work would be to consider an external magnetic field with an intrinsic correlation time and explore its effects, as this refinement would more accurately emulate real experimental conditions.

\begin{acknowledgments}
Partial financial support has been received from Grants PID2021-122256NB-C21/C22 and PID2024-157493NB-C21/C22 funded by MICIU/AEI/10.13039/501100011033 and by “ERDF/EU”, and the María de Maeztu Program for units of Excellence in R\&D, grant CEX2021-001164-M. S.O-B. acknowledges support from the Spanish Ministry of Education and Professional Training under the grant FPU21/04997. A.C. thanks Bret Flanders for many useful discussions. 
\end{acknowledgments}

\bibliography{References}

\end{document}